**BMC Genomics**

RESEARCH ARTICLE    Open Access

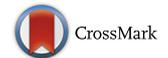

# Whole genome resequencing reveals diagnostic markers for investigating global migration and hybridization between minke whale species

Ketil Malde[1,5], Bjørghild B. Seliussen[1], María Quintela[1], Geir Dahle[1], Francois Besnier[1], Hans J. Skaug[1,4], Nils Øien[1], Hiroko K. Solvang[1], Tore Haug[3], Rasmus Skern-Mauritzen[1], Naohisa Kanda[2,7], Luis A. Pastene[2], Inge Jonassen[5] and Kevin A. Glover[1,6*]

## Abstract

**Background:** In the marine environment, where there are few absolute physical barriers, contemporary contact between previously isolated species can occur across great distances, and in some cases, may be inter-oceanic. An example of this can be seen in the minke whale species complex. Antarctic minke whales are genetically and morphologically distinct from the common minke found in the north Atlantic and Pacific oceans, and the two species are estimated to have been isolated from each other for 5 million years or more. Recent atypical migrations from the southern to the northern hemisphere have been documented and fertile hybrids and back-crossed individuals between both species have also been identified. However, it is not known whether this represents a contemporary event, potentially driven by ecosystem changes in the Antarctic, or a sporadic occurrence happening over an evolutionary time-scale. We successfully used whole genome resequencing to identify a panel of diagnostic SNPs which now enable us address this evolutionary question.

**Results:** A large number of SNPs displaying fixed or nearly fixed allele frequency differences among the minke whale species were identified from the sequence data. Five panels of putatively diagnostic markers were established on a genotyping platform for validation of allele frequencies; two panels (26 and 24 SNPs) separating the two species of minke whale, and three panels (22, 23, and 24 SNPs) differentiating the three subspecies of common minke whale. The panels were validated against a set of reference samples, demonstrating the ability to accurately identify back-crossed whales up to three generations.

**Conclusions:** This work has resulted in the development of a panel of novel diagnostic genetic markers to address inter-oceanic and global contact among the genetically isolated minke whale species and sub-species. These markers, including a globally relevant genetic reference data set for this species complex, are now openly available for researchers interested in identifying other potential whale hybrids in the world's oceans. The approach used here, combining whole genome resequencing and high-throughput genotyping, represents a universal approach to develop similar tools for other species and population complexes.

* Correspondence: kevin.glover@imr.no
[1]Institute of Marine Research, PO box 1870Nordnes, N-5817 Bergen, Norway
[6]Department of Biology, University of Bergen, N-5020 Bergen, Norway
Full list of author information is available at the end of the article

## Background

Anthropogenic factors contribute to increased rates of hybridization and introgression globally [1]. Moreover, the growing pace of species introductions and habitat alterations suggests that this phenomenon will become even more common in the future [2]. Global climate change is one of those anthropogenic factors that is expected to inevitably alter the prerequisites for many biota around the world [3] and a potential outcome will be a change in the niche separation of species with adjacent boundaries on both large and small scales. In marine ecosystems, which display few absolute physical barriers to migration and gene-flow, contemporary contact between previously isolated species can occur across great distances, and in some cases, may be inter-oceanic [4, 5]. Consequently, the identification of changes in species and population complexes is essential to document, quantify, and ultimately understand the potential long-term evolutionary consequences of ecosystem changes.

Minke whales represent a species complex in which it is possible to investigate how marine ecosystem changes can induce contemporary contact between genetically and geographically distinct species across the globe. Based upon morphological [6] and genetic data [7–9], this complex is thought to consist of two main species: the Antarctic minke whale (*Balaenoptera bonaerensis*) present in the southern hemisphere, and the common minke (*B. acutorostrata*), which is cosmopolitan. The common minke whale is thereafter divided into three allopatric sub-species: the North Atlantic (*B. acutorostrata acutorostrata*), the North Pacific (*B. a. scammoni*) and the dwarf common minke whale (*B. a.* unnamed sub-species) in the southern oceans [10]. Analyses of mtDNA data indicate that the two species may have been established from a separation in the southern hemisphere approximately 5 million years ago and that the sub-species diverged from each other approximately 1.5 million years ago [8].

Norway maintains an individual-based DNA register (NMDR) to enforce domestic regulation and compliance with the commercial harvest of the Atlantic minke whale *B. a. acutorostrata* conducted in the NE Atlantic [11]. This register contains genetic data corresponding to ten microsatellites and mtDNA for the whales harvested during the period 1996 till present, together with biometric information and the geographic position of the captures. In 1996, an Antarctic minke whale was captured in the NE Arctic [4], which represents the first documentation of this species north of the Equator, and serves as an example of a long distance and inter-oceanic migration. Approximately one decade later, in 2007, the first hybrid between Antarctic and Atlantic minke whales was identified, like the previous migrant this specimen was also captured in the NE Arctic [4]. In 2010, a second hybrid between these two species was also identified in the NE Arctic [7]. Significantly, this second hybrid was pregnant, and the genetic analyses of her normally developed fetus indicated that it had been sired by a common minke whale [7]. These observations confirm reproductive compatibility of hybrids between the Antarctic and Atlantic species of minke whale. Occasionally, routine genotyping with the standard panel of microsatellite markers used in the NMDR reveals specimens displaying atypical genetic profiles, which could suggest mixed ancestry. Although several of the microsatellite loci used are either fully or partially diagnostic among minke whale species and sub-species [7], the markers do not give sufficient statistical power to resolve the ancestry of these specimens.

The above observations (1996 migrant, 2007 hybrid, 2010 pregnant hybrid, 2010 back-crossed fetus, in addition to several unidentified but possibly back-crossed whales) provide a compelling time-line sequence of events. Documented changes in the Antarctic ecosystem [12] and changes in the energy storage and diet opportunities of Antarctic minke whales [13, 14] have led to speculation that Antarctic minke whales may be undertaking contemporary migrations out of its native distribution in search of better feeding opportunities in response to ecological changes [7]. While the microsatellite loci upon which the NMDR are based provide considerable statistical power to identify migrants and F1 hybrids [7], accurate characterization of individuals to various categories of hybrids and back-cross variants becomes increasingly challenging with subsequent generations. In order to investigate and follow-up the time-line of genetic contact between the minke whale species and sub-species as reported above, an improved set of markers is therefore necessary.

Simulations have demonstrated that 50 or more fully diagnostic markers (i.e., loci fixed to different alleles) are required to accurately identify F2 and F3 hybrids and multiple-generation back-crossed individuals [15, 16]. Recently, genomic resources for minke whales have become available [17, 18], and resequencing approaches to identify genetic markers [19] and bioinformatic tools to identify such loci have also become readily available for non-model species. Here, we used a combination of whole genome resequencing, SNP identification pipelines, and a high throughput genotyping platform to identify and validate a set of species diagnostic SNPs that can be used to provide accurate identification of hybrids and back-crossed minke whales globally.

## Methods

### Samples

The samples used in the present study originate from two sources. DNA from B. a. acutorostrata was obtained from the NMDR repository, using specimens from year classes 2007 and 2010 caught in the north-east Atlantic. Samples of B. bonaerensis were obtained during the JARPA (Japanese Whale Research Program under Special Permit in the Antarctic) survey performed in the austral summer season 2004/05 in International Whaling Commission (IWC) Management Area VIW (170°–145°W), south of 60°S. Samples of B. a. scammoni were obtained during the JARPNII (Japanese Whale Research Program under Special Permit in the North Pacific-Phase II) survey performed in summer 2006 in the western North Pacific between the Japanese Pacific coast and 170°E, and between 35° and 50°N approximately. Samples of B a. unnamed sub-species (dwarf minke whale) were obtained during the JARPA surveys performed in the austral summer seasons 1987/88–1992/93 in IWC Management Areas IV (70°–130°E) and V (130°E–170°W), south of 60°S. All samples existed prior to this study.

Table 1 gives an overview of the samples, classifying them into three partially overlapping categories: 1. Samples of all four species and subspecies that were selected for whole genome resequencing to identify putative species-diagnostic SNPs. 2. Samples of three of the four species and subspecies that were selected for high throughput genotyping to validate the SNPs and estimate allele frequencies for each species and subspecies. These results establish the genetic baseline, against which identification of "unknown" whales were performed. 3. Samples of the previously identified inter-oceanic migrant, F1 hybrids, and back-crossed fetus, as well as five individuals with slightly abnormal microsatellite profiles (all previously undescribed and originating from NMDR).

### Sample preparation

DNA from approximately 100 individuals from each of the three of the groups (not B. a. unnamed sub-species) was isolated using phenol-chloroform extraction. Isolated DNA was checked for size on agarose gels and for purity on a Nanodrop spectrophotometer, and was quantified using a Qubit Quant-iT kit from Invitrogen. From each of the three groups, 36 samples considered to be of the highest quality (size >10,000 bp, $Abs_{260}/Abs_{280}$ = 1.8–2 and $Abs_{260}/Abs_{230}$ = 1.8–2.4) were evenly pooled into two biological replicates each for the three groups.

The B. a. acutorostrata and the B. a. scammoni pools were purified with Genomic DNA Clean & Concentrator (DCC) kit from Zymo research, whereas the B. bonaerensis pools were purified by standard precipitation. The pools were subsequently checked for size, purity and quantity by the same method as the individual samples.

### Sequencing of pooled DNA samples

All subsequent laboratory work was performed at the Norwegian Sequencing Centre (NSC). Library construction was done using Illumina TruSeq adapter ligation. The DNA was fragmented to a target size of 300 bp, and sequenced using an Illumina HiSeq 2000 instrument, producing 2×101 bp paired-end reads. Each pool was sequenced using one Illumina lane and resulted in a pair of FASTQ files ranging from 45 to 50 GB in size (Table 2).

### Dwarf minke whale sequencing

Genomic DNA was extracted from approximately 0.05 g of the outer epidermal layer of the skin tissue of 11 dwarf minke whales using phenol–chloroform

**Table 1** Samples used in the present study

| Sample | Total | Sequenced | Genotyped |
|---|---|---|---|
| Samples for identification of markers, creation and validation of the genetic baseline | | | |
| B. bonaerensis | 95 | 36 | 95 |
| B. a. acutorostrata | 127 | 36 | 127 |
| B. a. scammoni | 95 | 36 | 95 |
| B. acutorostrata "dwarf" | 15 | 11 | 11[a] |
| Samples to be identified against the established genetic baseline | | | |
| ID 901016 (migrant, 1996) | 1 | 0 | 1 |
| ID 701550 (hybrid, 2007) [4] | 1 | 0 | 1 |
| ID 1001065 + 1001065_fetus (pregnant hybrid + fetus, 2010) [7] | 2 | 0 | 2 |
| ID 9601017, 1101069, 1101158, 1101205, 1401130 (Outlier/abnormal individuals from NMDR, 2009–2014)[b] | 5 | 0 | 5 |

[a]Dwarfs were sequenced in individually tagged lanes and their genotypes were inferred from the sequence data
[b]Previously unpublished, extracted from NMDR based on visual inspection of microsatellites

**Table 2** Sequencing results, showing the number of sequences produced for each pool, and the result of mapping them to the reference genome sequence

| Pool | Species | Individuals (N) | Reads (millions) | Mapped (millions) | Coverage (%) |
|---|---|---|---|---|---|
| AN1 | B. bonaerensis | 18 | 412.5 | 365.9 | 12.320 |
| AN2 | | 18 | 366.5 | 328.0 | 11.044 |
| AT1 | B. a. acutorostrata | 18 | 386.4 | 352.7 | 11.873 |
| AT2 | | 18 | 389.1 | 352.0 | 11.852 |
| PA1 | B. a. scammoni | 18 | 384.8 | 349.6 | 11.770 |
| PA2 | | 18 | 380.8 | 345.0 | 11.618 |
| DW | B. acutorostrata "dwarf" | 11 | 332.6 | 306.4 | 7.150 |

extraction. Extracted DNA was stored in TE buffer (10 mM Tris–HCl, 1 mM EDTA, pH 8.0).

DNA sequencing was performed by the Hokkaido System Science Co., Ltd (Sapporo, Japan). The DNA library was prepared with the TruSeq Nano DNA LT Sample Prep Kit (Illumina), following the manufacturer's standard protocol. Extracted DNA was fragmented into c.a. 350 bps, and fragments with other sizes were removed using the provided sample purification beads. The adaptor sequences were ligated to each end of the fragments. The resulting DNA library was sequenced using paired-end 100-bp reads on one lane of Illumina HiSeq 2500 system (Illumina Inc. USA). Bcl2fastq v1.8.3 (Illumina, San Diego, CA) was used to demultiplex the data into individual samples based on the indexes used during the library preparation.

### Mapping and SNP prediction

The *B. a. scammoni* genome reference sequence [18] was downloaded from the bioftp.org website, and the sequencing reads were mapped to this reference, using BWA version 0.7.5a [20] using the "mem" mapping method with default parameters, and samtools version 0.1.19 [21]. The mapped sequences were processed using *samtools mpileup* with the -B options, and then *varan* [22] was used with the -e option to tally the observed alleles and calculate the estimated information value (ESIV) for each locus in the genome.

### SNP candidate selection strategy

For a SNP to be highly informative for population or species identification, the difference in allele frequencies should be as large as possible. Ideally, the SNP should be fully diagnostic, i.e., fixed to different alleles in the different groups. In addition, SNPs should be independent of each other, avoiding correlation caused by genetic linkage. As exact chromosome location is not available, we depend on scaffold length to ensure maximum genomic distance between SNPs. Finally, for practical reasons, SNPs should be taken from relatively stable parts of the genome, avoiding repeats and highly polymorphic regions, since this complicates both mapping and design of primers to produce high through-put genotyping assays downstream.

In order to satisfy these criteria, we used the following strategy to select putatively species-diagnostic SNPs from the genome:

1. Select SNPs from a set of the longest genomic scaffolds.
2. Filter candidates on coverage, retaining only SNPs where coverage is within two standard deviations of the mean.
3. Select remaining candidates based on observed allele frequency difference and confidence, using the expected SNP information value (ESIV) [22].

### *B. acutorostrata* and *B. bonaerensis* species identification

Based on the approach detailed above, the two candidate SNP with loci with highest ESIV scores were selected from each of the 50 longest genome scaffolds. These candidates were processed with the MassARRAY® Typer 4.0 Assay Designer (Agena Bioscience) software to produce SNP primer multiplexes, suitable for the MassARRAY® genotyping platform (Agena Bioscience). In order to maintain maximal independence, when two loci from the same contig were produced, the locus with the lower confidence score was removed, and the multiplex was re-plexed in order to identify another suitable locus as a replacement.

### *B. acutorostrata* subspecies resolution

In order to develop markers to distinguish between the three *acutorostrata* subspecies (Atlantic, Pacific, and dwarf), a set of 100 candidate loci were selected for each subspecies, again selecting the two highest scoring loci from the 50 longest contigs. For sorting the candidate loci, the average of the ESIV scores between the subspecies and the two other groups was used.

### Primer design

The MassARRAY® Typer 4.0 Assay Design software (Agena Bioscience) was used to produce all SNP assays. Using the default values in the software (e.g., amplicon length 80–120 bp, and extension primer length 17–28 bp), and from the results, we selected 2 multiplex assays for the *acutorostrata* versus *bonaerensis* comparison (WP1 and WP2, with 26 and 28 SNPs respectively), and one multiplex for each of the three sub-species comparisons (WP3-5, with 23, 28 and 26 SNPs, see Table 3 for details).

### SNP genotyping and validation of allele frequencies

SNP genotyping was performed on a MassARRAY® Typer 4.0 Analyser (Agena Bioscience) at the molecular genetics laboratory at the Institute of Marine Research in Bergen. This platform is based on PCR amplification of the different SNPs, and each SNP in all multiplexes is detected in a high voltage vacuum, resulting in minute difference resolution, i.e. detection of two different fragments where only one nucleotide has been replaced. All data were analyzed and scored independently by two persons prior to exporting data to further analysis.

### Evaluation of the validated SNPs for species and sub-species identification

In order to evaluate the statistical effectiveness of the SNPs identified, two complimentary approaches were used. The first method is based on the observation that with fully diagnostic SNPs, only hybrids will display heterozygote loci, and if we consider only back-crosses into a homogenous, non-hybrid population, the number of heterozygotes will follow a binomial distribution. The second method performs a simulation using the observed allele frequencies in the species and sub-species, to estimate allele frequencies within the different hybrid and back-cross categories pair-wise, and thereafter examines the data using standard population genetics and genetic assignment tools (thus assigning each specimen into pre-defined categories). Both methods are described in detail below.

**Table 3** Marker panel genotyping results. Each assay contains between 22 and 26 usable markers, and 15 to 24 appear to be fully diagnostic – viz., we have not observed any presence of minor alleles in the two subspecies or populations. Except for the dwarf subspecies, where the number of specimens is small, a 95% confidence upper bound for the minor allele frequency of an apparently fully diagnostic marker is between 3 and 4%

| Panel | Population 1 | Population 2 | No markers used | "Fully diagnostic" markers |
| --- | --- | --- | --- | --- |
| WP1 | B. bonaerensis | B. acutorostrata | 26 | 24 |
| WP2 | B. bonaerensis | B. acutorostrata | 24 | 16 |
| WP3 | B. a. acutorostrata | Dwarf | 23 | 22 |
| WP4 | B. a. acutorostrata | B. a. scammoni | 22 | 15 |
| WP5 | B. a. scammoni | Dwarf | 24 | 21 |

First, we assessed classification from SNP markers based on heterozygote counts. We considered three scenarios: 1. assuming a panel of fully diagnostic markers without overlapping allele frequencies between the species, 2. assuming a panel of partly diagnostic markers, where the nearly-diagnostic alleles occur at small frequencies in the alternative population (both ways), and 3. assuming a panel of markers fully diagnostic in one direction, that is, where introgression migrants may have a small frequency, but where the loci are fixed (no minor alleles) in the native population. The latter case appears to be particularly relevant for *B. bonaerensis* introgressing into the Atlantic *B. acutorostrata* population. In this approach, we assumed that the identified markers were independent (i.e., there is no genetic-linkage), and treated the problem as a binomial sampling case. The details for these classifications are provided in Additional file 1.

Second, we used the population genetic and assignment approach based upon allele frequencies of whales in pre-defined categories (i.e., F1 hybrid, back-cross etc.). In short, this method involved producing *in silico* hybrids and various back-crossed whale combinations between *B. acutorostrata* and *B. bonaerensis* using allele frequency data from the genotyping results. Back-crossed whales were developed in both directions. Thereafter, the allele frequencies of the *in silico* developed hybrid and back-cross variants were compared against each other. First, a custom R script [23] was used to simulate the putative genotype of hybrid offspring, based on the known genotypes of two pools of parents. The initial data consisted of the genotypes of 95 Antarctic (*Balaenoptera bonaerensis*) and 95 Atlantic (*B. a. acutorostrata*) for 50 of the developed SNP markers. The first set of simulations was used to generate 500 F1 hybrids between Atlantic and Antarctic minke whales. Subsequently, the hybrids were iteratively crossed back with the original species to obtain reciprocal back-crosses as follows (Fig. 1): The F1 hybrids were crossed back with the Atlantic population to generate a first generation back-cross of Antarctic in Atlantic (F2A). Reciprocally, the F1 hybrids were crossed back with the Antarctic population to generate a first generation back-cross of Atlantic in Antarctic (F2B). Then, individuals from F2A were crossed with the Atlantic population to obtain a second generation back-cross of Antarctic in Atlantic (F3A), and individuals from F2B were crossed with the Antarctic population to obtain a second generation back-cross of Atlantic in Antarctic (F3B). The procedure was repeated until the fourth generation of back-cross (F5A and F5B). Thereafter, the accuracy of the assignment of simulated hybrids to their potential source was calculated by testing 457 hybrids against the baseline of 1101 individuals distributed in eleven groups (i.e., two pure species and the nine different hybrid and backcross variants). This analysis was conducted with the program GeneClass 2 [24] using the Rannala & Mountain [25] method of computation as well as with the conditional maximum likelihood based approach implemented in the software ONCOR (S.Kalinowski, http://www.montana.edu/kalinowski/Software/ONCOR.htm). We also used the Bayesian model-based clustering algorithms implemented in STRUCTURE v. 2.3.4 [26] to identify genetic clusters in the 1101 individual baseline data set under a model assuming admixture and correlated allele frequencies both with and without using population information. In each case, ten runs with a burn-in period consisting of 100,000 replications and a run length of 1,000,000 Markov chain Monte Carlo (MCMC) iterations were performed for a number of clusters ranging from $K = 1$ to $K = 6$. Runs were automatized with the program Parallel Structure [27] that controls the program STRUCTURE and distributes jobs between parallel processors to significantly speed up the analysis time. We then used STRUCTURE Harvester to calculate the Evanno ad hoc statistic ΔK, which is based on the rate of change of the 'estimated likelihood' between successive K values [28]. Each genotype was then assigned to the inferred clusters based of the individual proportion of membership ($q_i$) and its confidence interval (90% CI). Q-values and their corresponding 90% CI were averaged for each of the populations and plotted. STRUCTURE runs for the most likely K were averaged with CLUMPP version 1.1.1 [29] using the LargeKGreedy algorithm and the G' pairwise matrix similarity statistics, and results were graphically displayed using barplots.

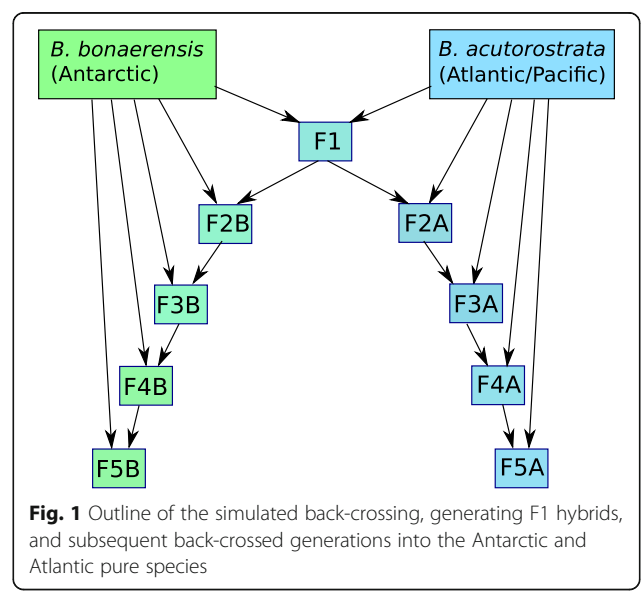

**Fig. 1** Outline of the simulated back-crossing, generating F1 hybrids, and subsequent back-crossed generations into the Antarctic and Atlantic pure species

## Results and discussion
### Summary of the identified markers

As described above, five panels of putatively diagnostic SNPs were produced. These SNPs were thereafter validated by genotyping the described samples (Table 1). An overview of this validation is given in Table 3 (for full details, see Additional file 2). Since no DNA samples from dwarf minke were available for SNP genotyping, genotypes for these 11 individuals were estimated from the individually-tagged sequencing data. The sequencing data from dwarf minke showed none of the SNP minor alleles found in the other groups. Concluding that the SNPs are diagnostic for this group is premature, however, since using the same data for SNP prediction and for testing incurs a selection bias. In addition, the data is based on fewer individuals, further reducing confidence in such a conclusion. Thus, the likelihood for the existence of undetected minor alleles (and thus markers that are not fully diagnostic) is much higher for dwarf minke than for the other species and subspecies.

Some of the SNPs designed to separate *B. bonaerensis* and *B. acutorostrata* (panels WP1 and WP2) were not completely diagnostic in the samples used for validation. In all cases, this was caused by a very low presence (estimated MAFs ranging from 0.6 to 4.7%) of the *B. acutorostrata* alleles in the *B. bonaerensis* specimens. A greater genetic diversity within *B. bonaerensis* is to be expected, due to the greater population size of this species, and the result is consistent with previously reported results based on microsatellites [7].

Some of the SNPs failed to amplify in some of the specimens, and thus the actual number of individuals genotyped varied per SNP and per species. The number of individuals successfully genotyped per SNP ranged from 76 to 95 for *B. a. scammoni* samples, 110 to 117 for *B. a. acutorostrata*, and 81 to 94 for *B. bonaerensis*. Using the Agresti-Coull approximation [30], we found that even for SNPs where no minor allele was observed, the 95% confidence interval for the minor allele frequency was from 0 to between 3.3 and 4.1%.

The dwarf minke is considered to be a member of the common minke species [31, 32]. Examining marker panel WP4, we found that the dwarf data matched the *B. a. acutorostrata* allele for 12 SNPs, the *B. a. scammoni* allele for 8 SNPs, and for two of the SNPs we found both alleles present. The separation is here by SNP, not by individual, so this appears to be the result of the dwarf minke constituting a separate sub-species with genetic similarities with both the *B. a. acutorostrata* and the *B. a. scammoni* sub-species.

### Identification of hybrids and back-crossed whales from heterozygote counts

From the observed number of heterozygotes, the ancestry of individuals were estimated using maximum likelihood. Table 4 summarizes the results of this analysis, further details are provided in Additional files 1 and 3. Under realistic minor allele frequencies, F1 and F2 hybrids are correctly identified in almost all (>95%) cases. While F3 and F4 generations will be correctly identified in most of the cases (see Table 4 for details), they will occasionally be misidentified, mostly as a more remote back cross. The accuracy of identification depends on assumptions about the minor allele frequencies for the SNPs, as can be seen from Table 3, the worst case presented here (i.e., a MAF of 0.05) is likely to be overly pessimistic.

Although maximum likelihood assignment is a simple approach, it ignores the prior probability of observing the different generations of hybrids. An argument could be made that a Bayesian assignment would be more accurate, but this is conditioned on the availability of good

**Table 4** Classification intervals, giving the range of heterozygote marker counts leading to Fn classification under different assumptions of allele fixedness, and the probability of the classification of the specific back cross generation being correct, using a maximum likelihood model

|  | Diagnostic markers | | | Introgressing MAF = 0.05 | | | MAF = 0.05 | | |
| --- | --- | --- | --- | --- | --- | --- | --- | --- | --- |
|  | from | to | prob. correct | from | to | prob. correct | from | to | prob. correct |
| F1 | 36 | 50 | 1.000 | 36 | 50 | 1.000 | 36 | 50 | 1.000 |
| F2 | 19 | 35 | 0.966 | 18 | 35 | 0.963 | 20 | 35 | 0.940 |
| F3 | 10 | 18 | 0.808 | 9 | 17 | 0.836 | 13 | 19 | 0.687 |
| F4 | 5 | 9 | 0.677 | 5 | 8 | 0.590 | 9 | 12 | 0.501 |
| F5 | 3 | 4 | 0.411 | 3 | 4 | 0.402 | 7 | 8 | 0.306 |
| F6 | 2 | 2 | 0.261 | 2 | 2 | 0.254 | 6 | 6 | 0.171 |
| F7 | 1 | 1 | 0.361 | 1 | 1 | 0.357 |  |  |  |
| Native | 0 | 0 | 1.000 | 0 | 0 | 1.000 | 0 | 4 | 0.479 |

Note that with minor allele frequency of 0.05, there is no number of heterozygote markers where the F7 generation is most likely to produce, and thus we will never classify a specimen as F7. The same holds for F8 and above under all assumptions of minor allele frequencies.

priors. Assuming a constant population and no fitness advantage or disadvantage for the mixed ancestry individuals, we would expect a doubling in the number of specimens for each generation - i.e., each migrant will be the parent of two F1 hybrids, who in turn are parent to four F2 hybrids, and so on. There is little evidence that supports the assumptions underlying this model. If migration is indeed driven by recent changes to climate and ecosystems, a typical generation time for minke whales of 15–20 years [18] means that we are unlikely

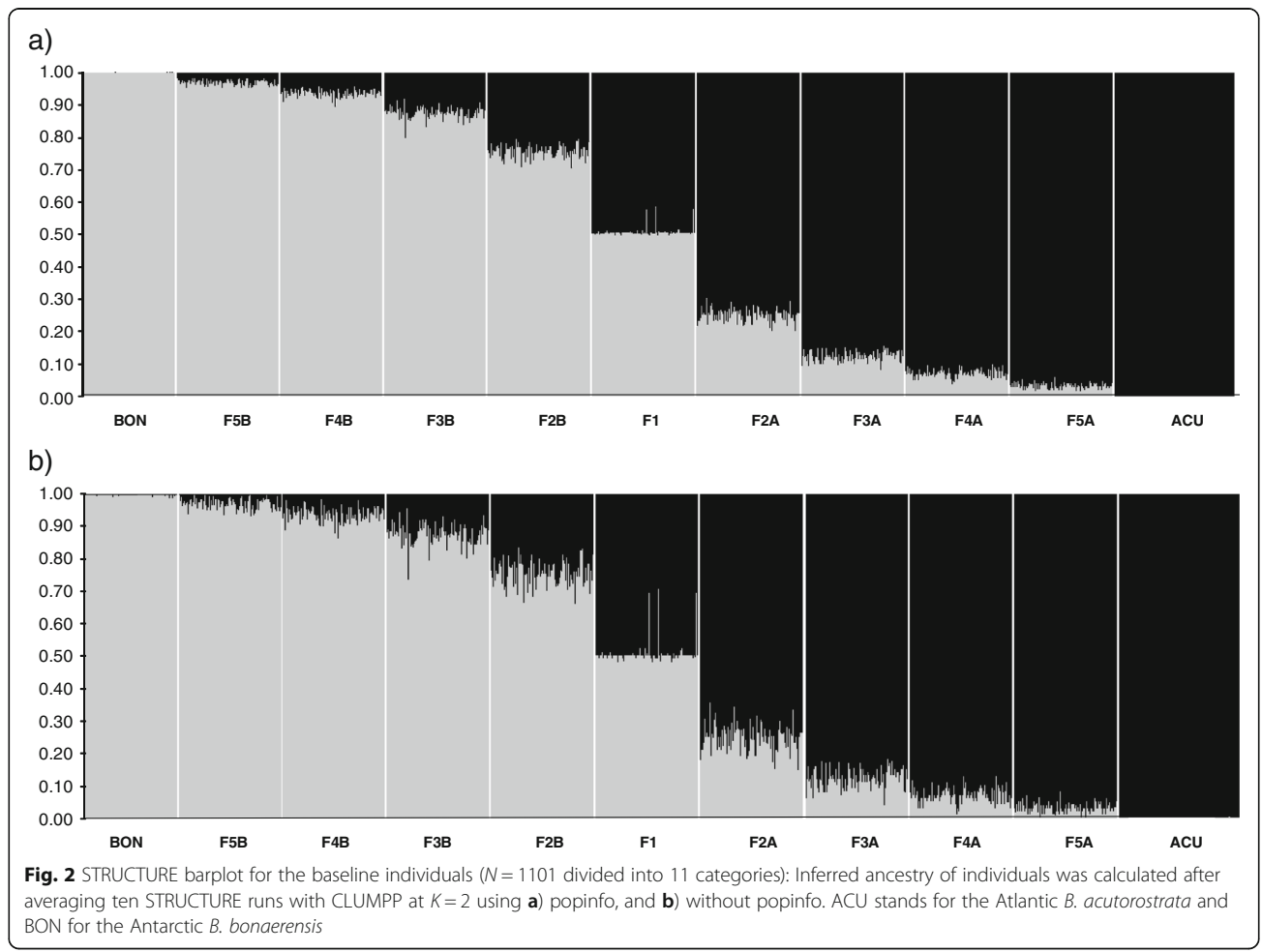

**Fig. 2** STRUCTURE barplot for the baseline individuals (N = 1101 divided into 11 categories): Inferred ancestry of individuals was calculated after averaging ten STRUCTURE runs with CLUMPP at K = 2 using **a**) popinfo, and **b**) without popinfo. ACU stands for the Atlantic *B. acutorostrata* and BON for the Antarctic *B. bonaerensis*

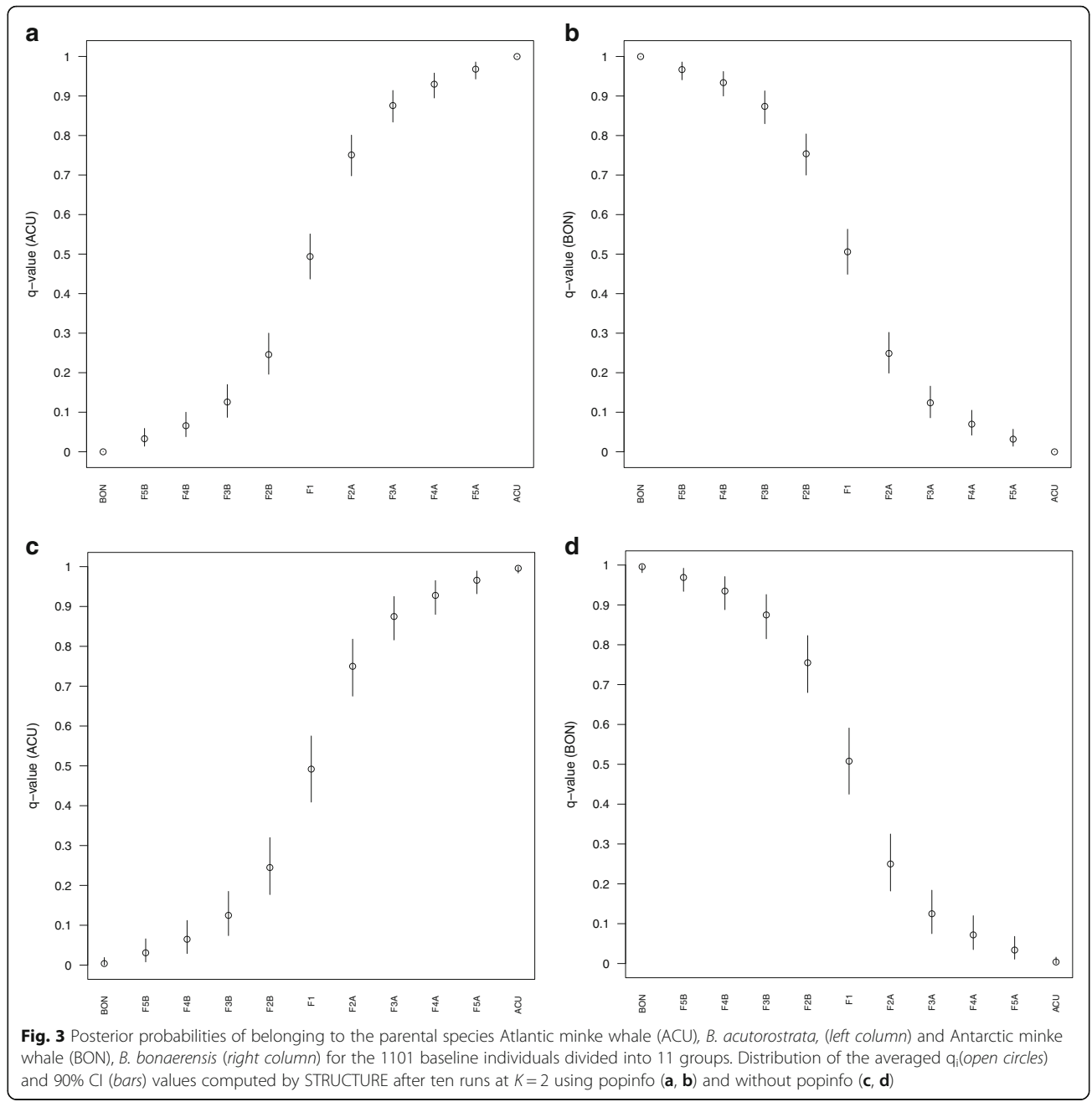

**Fig. 3** Posterior probabilities of belonging to the parental species Atlantic minke whale (ACU), *B. acutorostrata*, (*left column*) and Antarctic minke whale (BON), *B. bonaerensis* (*right column*) for the 1101 baseline individuals divided into 11 groups. Distribution of the averaged $q_i$ (*open circles*) and 90% CI (*bars*) values computed by STRUCTURE after ten runs at $K = 2$ using popinfo (**a**, **b**) and without popinfo (**c**, **d**)

to see hybrids further back than three or four generations.

Our current understanding of the minke whale indicates a relatively small number of *B. bonaerensis* migrants introgressing into *B. acutorostrata* [4, 7]. In this situation, hybrids are likely to continue to back-cross with *B. acutorostrata*, until only a small remnant of the introgressed genes remain in each individual. We have therefore emphasized this scenario in our analysis above.

If two hybrids should happen to interbreed, we expect to see homozygote markers specific to both ancestor populations (in our case, *B. bonaerensis* and *B. acutorostrata*). If this occurs at an early enough generation, it should be easy to detect, for example, for the offspring of two F1 hybrids (all markers heterozygote), we would expect 25% homozygote markers matching each ancestral population. Similar to the situation for identifying an F2 back-cross, we have a 99.8% chance of at least three homozygote markers from each ancestral population, thus revealing the mixed ancestry. For the offspring of two F2 back-crosses, we expect the marker fractions to be 0.5625 homozygote to the local population, 0.375 heterozygote, and 0.0625 foreign homozygote. In this case, a presence of foreign homozygote markers would reveal the mixed heritage, but in an expected 4% of the cases, no such markers are produced, and the individual would be misclassified as an F1 or F2 back-cross.

Table 5 Distribution of 457 simulated test individuals into the baseline of 1101 individuals. Correct assignment is depicted in boldface type

| Population | BON | F5B | F4B | F3B | F2B | F1 | F2A | F3A | F4A | F5A | ACU |
|---|---|---|---|---|---|---|---|---|---|---|---|
| F2B | | | | 1 | **50** | | | | | | |
| F3B | | | 9 | **42** | | | | | | | |
| F4B | | 23 | **24** | 4 | | | | | | | |
| F5B | 9 | **35** | 6 | 1 | | | | | | | |
| F1 | | | | | | **51** | | | | | |
| F2A | | | | | | 1 | **48** | 2 | | | |
| F3A | | | | | | | 4 | **35** | 10 | 1 | |
| F4A | | | | | | | | 8 | **22** | 20 | |
| F5A | | | | | | | | 1 | 11 | **33** | 6 |

### Identification of hybrids and back-crossed whales using genetic assignment

The program STRUCTURE distinctly showed two main clusters between the simulated hybrids and back-crosses between *B bonaerensis* and *B. a. acutorostrata* (Evanno's ΔK = 13,893). A clear gradient of admixture between these two species was apparent from these in silico established crosses (Fig. 2a, b). The degree of admixture using popinfo prior was notably lower than the one found when this option was disabled. The 90% confidence intervals around the averaged $q_i$ per population were narrower for the popinfo option (Fig. 3). Hence, under popinfo, the only simulated hybrids that were impossible to differentiate due to overlapping confidence intervals were F3 from F4 and F4 from F5, whereas all the remaining combinations were distinguishable (Fig. 3a, b). Without popinfo, some overlaps add to the aforementioned ones: BON vs. F5B, ACU vs. F5A, and F2 vs. F3 (Fig. 3c, d). The assignment of individuals to pure species or hybrid categories depends upon a trade-off between the accuracy and the efficiency in the assignment. When using STRUCTURE, the relatively stringent posterior probability threshold of 0.90 minimizes the number of misidentified pure species individuals while maximizing the efficiency of assigning hybrids (see [33]). Thus, the highest efficiency, i.e., the highest proportion of correctly classified purebreds and hybrids, is achieved with a threshold q-value of 0.10. However, when accuracy, i.e., the proportion of true purebreds or true hybrids in each the purebred or hybrid groups, was considered in addition to efficiency, the threshold q-value of 0.20 resulted in 6% higher performance. Following this rationale, the threshold of 0.10 < q < 0.90 is the most widely utilized in the literature dedicated to the assessment of hybridization (e.g. [34–36]). However, in our case this threshold is too wide to allow the differentiation between pure species (ACU, BON) and F3 and F4 backcrosses as, even with no popinfo, our threshold for hybrids would be 0.016 < q < 0.989.

GeneClass showed that all F1 individuals were correctly assigned whereas the percentage of correct assignment for F2 ranged between 94 and 98% and between 70 and 82% for F3 (Table 5). Incorrectly assigned individuals were always directed into their neighboring categories.

Table 6 Summary of classification of migrant and hybrid specimens according different methods: the original classification using microsatellites, the classification using SNP-specific markers from the *B. bonaerensis* (BON) vs. *B. a. acutorostrata* (ACU) specific assays WP1 and WP2, the classification using SNP without 5% MAF, the direct assignment using STRUCTURE and ONCOR against a simulated SNP-based baseline. Column "ID" gives the internal NMDR identification numbers

| ID | Classif microsatellites | Classif SNP | Classif SNP w/ 5% MAF | STRUCTURE classif against simulated SNP-based baseline | GeneClass direct assignment against simulated SNP-based baseline (and score, %) | ONCOR direct assignment against simulated SNP-based baseline (and probability) |
|---|---|---|---|---|---|---|
| 9601017 [4] | BON | BON | BON | BON | BON (93.86) | BON (1.000) |
| 701550 [4] | F1 | F1 | F1 | F1 | F1 (100) | F1 (1.000) |
| 901016[a] | ACU (outlier) | ACU | ACU | ACU | ACU (93.74) | ACU (1.000) |
| 1001065 [7] | F1 | F1 | F1 | F1 | F1 (100) | F1 (1.000) |
| 1001065_fetus [7] | F2 (F1 + ACU) | F2A | F2A | F2A | F2A (99.92) | F2A (0.999) |
| 1101069[a] | ACU (outlier) | ACU | ACU | ACU | ACU (93.90) | ACU (1.000) |
| 1101158[a] | ACU (outlier) | F2A, most likely | F3A/F2A | F2A | F2A (76.89) | F2A (1.000) |
| 1101205[a] | ACU (outlier) | ACU | ACU | ACU | ACU (93.90) | ACU (1.000) |
| 1401130[a] | ACU (outlier) | ACU | ACU | ACU | ACU (93.74) | ACU (1.000) |

[a]Previously unpublished, extracted from NMDR based on visual inspection of microsatellites

### Identification of the hybrids and back-crossed whales in the NMDR

We used the diagnostic SNPs and genetic baselines described above to identify all the hybrids, back-crossed whales and unidentified whales with a deviating microsatellite DNA profile for *B. A. acutorostrata*. All the individuals with a strong classification from their microsatellite profile (9601017 (migrant), 701550 (first F1), 1001065 (second F1) and 1001065-fetus (first F2 detected)) had their classification confirmed by the diagnostic SNPs developed here (Table 6). Among the five outlying Atlantic individuals (901016, 1101069, 1101158, 1101205 and 1401130), one (1101158) was classified as an F2 or possible F3, while the rest were classified as pure Atlantic individuals.

## Conclusions

In the present study, we have used a combination of whole-genome re-sequencing, SNP detection pipelines, high through-put genotyping, and statistical methods to identify, validate and demonstrate the identification power of a suite of species-diagnostic SNPs within the minke whale species and sub-species complex. Most of the putatively diagnostic markers, identified from sequencing, were demonstrated to be fully or nearly diagnostic among the various species and sub-species combinations. Collectively, these markers permit highly accurate identification of species, sub-species, F1 hybrids, and back-crossed whales in this species-complex. It is also likely that additional mining of the sequence data developed within this project will provide identify diagnostic markers if required. Given the fact that minke whales have shown atypical inter-oceanic migrations in recent years [4] and that hybrids between these species appear to be fertile and may therefore breed across currently accepted species boundaries [7], the SNPs developed here, in addition to the available reference data sets that we have made publicly available, will provide the international community with the ability to investigate the movements and genetics of this species complex in all regions of the globe.

## Additional files

**Additional file 1:** Probabilities for heterozygote genetic markers in hybrids. (PDF 29 kb)

**Additional file 2:** Genotyping results. (XLS 1294 kb)

**Additional file 3:** Classification of genotyped samples. (ODS 136 kb)

### Acknowledgements
We would like to acknowledge the assistance of Lotta Lindblom and Anne-Grete Sørvik from the Institute of Marine Research for organizing tissue samples within the NMDR and contributing to the genotyping respectively. This study was financed by the Norwegian Ministry of Industry and Fisheries, although they had no participation in the design of the study nor in the interpretation of the results.

### Availability of data and materials
Sequencing data is available from the Sequencing Read Archive as BioProject PRJNA335553.

### Authors' contributions
KAG conceived and managed the study. KM conducted all bioinformatics analysis, wrote the first draft of the manuscript, and coordinated mathematical simulations. RSM and NK organized sequencing. BBS and GD conducted genotyping and data quality checks. HJS, FB and MQ contributed to simulations. All authors, including those not specifically mentioned with tasks as above, contributed to the design of the study, interpretation of the results, and approved the final version of the manuscript.

### Competing interests
The authors declare that they have no competing interests.

### Consent for publication
Not applicable.

### Ethics approval and consent to participate
This study exclusively uses pre-existing tissue samples, and no live animals were involved.

### Author details
[1]Institute of Marine Research, PO box 1870Nordnes, N-5817 Bergen, Norway. [2]Institute of Cetacean Research, Toyomi-cho 4-5, Chuo-ku, Tokyo 104-0055, Japan. [3]Institute of Marine Research, PO box 6404N-9294 Tromsø, Norway. [4]Department of Mathematics, University of Bergen, N-5020 Bergen, Norway. [5]Department of Informatics, University of Bergen, N-5020 Bergen, Norway. [6]Department of Biology, University of Bergen, N-5020 Bergen, Norway. [7]Japan NUS Co., Ltd, Nishi-Shinjuku Kimuraya Bldg 5F, 7-5-25, Nishi-Shinjuku 160-0023, Japan.

### References
1. Rhymer JM, Simberloff D. Extinction by hybridization and introgression. Annu Rev Ecol Syst. 1996;27(1):83–109.
2. Allendorf FW, Luikart G. Conservation and the genetics of populations. Malden: Blackwell Pub; 2007: xix, 642 p.
3. Vitousek PM, D'antonio CM, Loope LL, Rejmanek M, Westbrook R. Introduced species: a significant component of human-caused global change. N Z J Ecol. 1997;21(1):1–16.
4. Glover KA, Kanda N, Haug T, Pastene LA, Øien N, Goto M, Seliussen BB, Skaug HJ. Migration of Antarctic minke whales to the Arctic. PLoS One. 2010;5(12):e15197.
5. Pomilla C, Rosenbaum HC. Against the current: an inter-oceanic whale migration event. Biol Lett. 2005;1(4):476–9.
6. Omura H. Ostelogical study of the minke whale from the Antarctic. Scientific Reports of the Whales Research Institute. 1975;27:1–36.
7. Glover KA, Kanda N, Haug T, Pastene LA, Øien N, Seliussen BB, Sørvik AGE, Skaug HJ. Hybrids between common and Antarctic minke whales are fertile and can back-cross. BMC Genet. 2013;14(1):25.
8. Pastene LA, Goto M, Kanda N, Zerbini AN, Kerem DAN, Watanabe K, Bessho Y, Hasegawa M, Nielsen R, Larsen F, et al. Radiation and speciation of pelagic organisms during periods of global warming: the case of the common minke whale, *Balaenoptera acutorostrata*. Mol Ecol. 2007;16(7):1481–95.
9. Wada S. Genetic distinction between two minke whale stocks in the Okhotsk Sea coast of Japan. In: Reports of the International Whaling Commission. vol. SC/43/Mi32; 1991.
10. Rice DW. Marine Mammals of the World. Systematics and Distribution, vol. 4. 1998.
11. Glover KA, Haug T, Øien N, Walløe L, Lindblom L, Seliussen BB, Skaug HJ. The Norwegian minke whale DNA register: a data base monitoring commercial harvest and trade of whale products. Fish Fish. 2012;13(3):313–32.
12. Atkinson A, Siegel V, Pakhomov E, Rothery P. Long-term decline in krill stock and increase in salps within the Southern Ocean. Nature. 2004; 432(7013):100–3.

13. Konishi K, Tamura T, Zenitani R, Bando T, Kato H, Walløe L. Decline in energy storage in the Antarctic minke whale (*Balaenoptera bonaerensis*) in the Southern Ocean. Polar Biol. 2008;31(12):1509–20.
14. Konishi K, Hakamada T, Kiwada H, Kitakado T, Walløe L. Decrease in stomach contents in the Antarctic minke whale (*Balaenoptera bonaerensis*) in the Southern Ocean. Polar Biol. 2014;37(2):205–15.
15. Fitzpatrick BM, Johnson JR, Kump DK, Shaffer HB, Smith JJ, Voss SR. Rapid fixation of non-native alleles revealed by genome-wide SNP analysis of hybrid tiger salamanders. BMC Evol Biol. 2009;9(1):176.
16. Fitzpatrick BM. Estimating ancestry and heterozygosity of hybrids using molecular markers. BMC Evol Biol. 2012;12(1):131.
17. Park JY, An Y-R, Kanda N, An C-M, An HS, Kang J-H, Kim EM, An D-H, Jung H, Joung M, et al. Cetaceans evolution: insights from the genome sequences of common minke whales. BMC Genomics. 2015;16(1):13.
18. Yim H-S, Cho YS, Guang X, Kang SG, Jeong J-Y, Cha S-S, Oh H-M, Lee J-H, Yang EC, Kwon KK, et al. Minke whale genome and aquatic adaptation in cetaceans. Nat Genet. 2014;46(1):88–92.
19. Kumar S, Banks TW, Cloutier S. SNP discovery through Next-Generation Sequencing and its applications. International Journal of Plant Genomics. 2012;2012:15.
20. Li H, Durbin R. Fast and accurate short read alignment with Burrows–Wheeler transform. Bioinformatics. 2009;25(14):1754–60.
21. Li H, Handsaker B, Wysoker A, Fennell T, Ruan J, Homer N, Marth G, Abecasis G, Durbin R, GPDP Subgroup. The Sequence Alignment/Map format and SAMtools. Bioinformatics. 2009;25(16):2078–9.
22. Malde K. Estimating the information value of polymorphic sites using pooled sequences. BMC Genomics. 2014;15 Suppl 6:S20.
23. Team RDC. R: a language and environment for statistical computing. Vienna: R Foundation for Statistical Computing; 2015.
24. Piry S, Alapetite A, Cornuet J-M, Paetkau D, Baudouin L, Estoup A. GENECLASS2: a software for genetic assignment and first-generation migrant detection. J Hered. 2004;95(6):536–9.
25. Rannala B, Mountain JL. Detecting immigration by using multilocus genotypes. Proc Natl Acad Sci U S A. 1997;94(17):9197–201.
26. Pritchard JK, Stephens M, Donnelly P. Inference of population structure using multilocus genotype data. Genetics. 2000;155(2):945–59.
27. Besnier F, Glover KA. ParallelStructure: a R package to distribute parallel runs of the population genetics program STRUCTURE on multi-core computers. PLoS One. 2013;8(7):e70651.
28. Evanno G, Regnaut S, Goudet J. Detecting the number of clusters of individuals using the software structure: a simulation study. Mol Ecol. 2005;14(8):2611–20.
29. Jakobsson M, Rosenberg NA. CLUMPP: a cluster matching and permutation program for dealing with label switching and multimodality in analysis of population structure. Bioinformatics. 2007;23(14):1801–6.
30. Agresti A, Coull BA. Approximate is better than "exact" for interval estimation of binomial proportions. Am Stat. 1998;52(2):119–26.
31. Arnold PW, Birtles RA, Dunstan A, Lukoschek V, Matthews M. Colour patterns of the dwarf minke whale *Balaenoptera acutorostrata* sensu lato: description, cladistic analysis and taxonomic implications. Memoirs of the Queensland Museum. 2005;51:277–307.
32. Pastene LA, Goto M, Kanda N. Progress in the development of stock structure hypotheses for western North Pacific common minke whales. SC/62/NPM12rev 2010.
33. Vähä J-P, Primmer CR. Efficiency of model-based Bayesian methods for detecting hybrid individuals under different hybridization scenarios and with different numbers of loci. Mol Ecol. 2006;15(1):63–72.
34. Winkler KA, Pamminger-Lahnsteiner B, WanzenbÖCk J, Weiss S. Hybridization and restricted gene flow between native and introduced stocks of Alpine whitefish (*Coregonus sp.*) across multiple environments. Mol Ecol. 2011;20(3):456–72.
35. Dierking J, Phelps L, Præbel K, Ramm G, Prigge E, Borcherding J, Brunke M, Eizaguirre C. Anthropogenic hybridization between endangered migratory and commercially harvested stationary whitefish taxa (*Coregonus spp.*). Evol Appl. 2014;7(9):1068–83.
36. Burgarella C, Lorenzo Z, Jabbour-Zahab R, Lumaret R, Guichoux E, Petit RJ, Soto A, Gil L. Detection of hybrids in nature: application to oaks (*Quercus suber* and *Q. ilex*). Heredity. 2009;102(5):442–52.